\documentclass[12pt,a4paper,final]{iopart}

\usepackage{iopams}
\usepackage{multirow}
\usepackage{stackengine}
\usepackage{graphicx}
\usepackage{cite}
\usepackage{subfig}

\usepackage[breaklinks=true,colorlinks=true,linkcolor=blue,urlcolor=blue,citecolor=blue]{hyperref}

\begin{document}

\title[Flexible superconducting Nb transmission lines]{Flexible superconducting Nb transmission lines on thin film polyimide for quantum computing applications}

\author{David B. Tuckerman$^{1}$, Michael C. Hamilton$^{2}$, David J. Reilly$^{3}$, Rujun Bai$^{2}$, George A. Hernandez$^{2}$, John M. Hornibrook$^{3}$, John A. Sellers$^{2}$ and Charles D. Ellis$^{2}$}
\address{$^1$Microsoft Research, Redmond, WA, 98052, USA}
\address{$^2$Alabama Micro/Nano Science and Technology Center, Electrical and Computer Engineering Department, Auburn University, Auburn, AL, 36849, USA}
\address{$^3$ARC Centre of Excellence for Engineered Quantum Systems and Station Q Sydney, School of Physics, The University of Sydney, Sydney, NSW 2006, Australia}

\begin{abstract}
We describe progress and initial results achieved towards the goal of developing integrated multi-conductor arrays of shielded controlled-impedance flexible superconducting transmission lines with ultra-miniature cross sections and wide bandwidths (dc to $>$10 GHz) over meter-scale lengths. Intended primarily for use in future scaled-up quantum computing systems, such flexible thin-film niobium/polyimide ribbon cables could provide a physically compact and ultra-low thermal conductance alternative to the rapidly increasing number of discrete coaxial cables that are currently used by quantum computing experimentalists to transmit signals between the several low-temperature stages (from $\sim$4 K down to $\sim$20 mK) of a dilution refrigerator. We have concluded that these structures are technically feasible to fabricate, and so far they have exhibited acceptable thermo-mechanical reliability. S-parameter results are presented for individual 2-metal layer Nb microstrip structures having 50 $\Omega$ characteristic impedance; lengths ranging from 50 to 550 mm were successfully fabricated. Solderable pads at the end terminations allowed testing using conventional rf connectors. Weakly coupled open-circuit microstrip resonators provided a sensitive measure of the overall transmission line loss as a function of frequency, temperature, and power. Two common microelectronic-grade polyimide dielectrics, one conventional and the other photo-definable (PI-2611 and HD-4100, respectively) were compared. Our most striking result, not previously reported to our knowledge, was that the dielectric loss tangents of both polyimides, over frequencies from 1 to 20 GHz, are remarkably low at deep cryogenic temperatures, typically 100$\times$ smaller than corresponding room temperature values. This enables fairly long-distance (meter-scale) transmission of microwave signals without excessive attenuation, and also permits usefully high rf power levels to be transmitted without creating excessive dielectric heating. We observed loss tangents as low as 2.2$\times$10$^{-5}$ at 20 mK, although losses increased somewhat at very low rf power levels, similar to the well-known behavior of amorphous inorganic dielectrics such as SiO$_2$. Our fabrication techniques could be extended to more complex structures such as multiconductor cables, embedded microstrip, 3-metal layer stripline or rectangular coax, and integrated attenuators and thermalization structures. 

\end{abstract}

\vspace{2pc}


\section{Introduction}
There is currently great interest in building quantum computers that could perform certain types of useful computations (e.g., quantum chemistry) faster and more cost-effectively than any existing digital computer technology. The error rates of certain types of qubit technologies have now improved to the point that it seems that such a feat of engineering may become possible in the next decade, albeit very challenging and costly\cite{Barends2014}. To be economically compelling, a quantum computer should probably contain the equivalent of at least several hundred sufficiently error-free qubits, or ``logical qubits''. Unfortunately, physically realizable qubits (``physical qubits'') are prone to occasional errors due to relaxation and/or phase decoherence, quantum logic gate operations are not perfectly accurate, and the measurement of quantum states can return erroneous results as well as cause back-action on those states. Quantum error correction has recently been experimentally demonstrated to extend the lifetime of quantum information \cite{Reed2012,Kelly2015,Chow2015,Ofek2016}, but to achieve usefully long coherence times, the physical ``overhead'' (ratio of the number of physical qubits required to emulate a given number of logical qubits) is extraordinarily large: factors of thousands or even millions may well be needed, depending on one's assumptions about error rates \cite{Fowler2012}. Recent progress towards creating ``topologically protected'' solid-state qubits (e.g., based on Majorana zero modes) offers the hope that qubit error rates could be made exponentially small, producing a near-perfect ``quantum memory" \cite{Mourik2012,Albrecht2015}. Schemes have also been proposed by which certain quantum gate operations (the Clifford group) could also be implemented in a topologically protected manner (specifically, by ``braiding'' Majorana zero modes, which are predicted to behave as non-Abelian anyons) \cite{Hyart2013}. However, the Clifford gates are insufficient by themselves to realize a universal quantum computer, and therefore, some substantial level of physical overhead for quantum error correction (e.g., to perform ``state distillation'') would still be required \cite{Bravyi2005}.

It seems, therefore, that even in the most optimistic scenario wherein extremely low physical error rates are achieved, a useful quantum computer will require many thousands of physical qubits, and less optimistic scenarios will require millions. This is now well understood by qubit device researchers, and there have been a number of recent proposals as to how one might architect systems that are scalable to large numbers of qubits \cite{Brecht2015,Hornibrook2015}. Most of these proposals deal with how qubits could be integrated and interconnected in 2-dimensional arrays, using compact wafer-like configurations.  There has been limited attention given to the physical problem of how to transmit a large number of control signals to those qubits in order to drive simultaneous gate operations, and how to receive a large number of simultaneous measurement signals from the qubit array, particularly when these signals need to span a large temperature difference (e.g., different stages of a dilution refrigerator). Our work is intended to address this specific interconnection problem.

With the exception of optically excited qubits that can operate at room temperature, most scalable solid-state qubit technologies (e.g., superconducting qubits, spin qubits) are principally controlled by precisely shaped microwave pulses; these frequencies are most commonly in the 5 to 15 GHz range. These pulses may be used to drive transitions between two states of a qubit (e.g., between the primary $\vert0\rangle$ and $\vert1\rangle$ states), to create ``cross-resonance'' conditions \cite{Chow2013}, or to induce sideband transitions\cite{Leek2009}. Microwaves are also commonly used to measure the state of a resonator-coupled qubit (i.e., ``dispersive readout'')\cite{Wallraff2005}. Furthermore, microwaves are also used as pump tones for low-noise superconducting parametric amplifiers in conjunction with qubit readout \cite{Macklin2015}.

The use of 5 to 15 GHz control signals necessarily implies that the qubits be cooled to temperatures in the 10s of mK to avoid thermal population of the $\vert1\rangle$ state; the qubits also need to be well-shielded from higher-temperature blackbody radiation, which could act as electromagnetic interference or sources of nonequilibrium quasiparticles \cite{Corcoles2011,Barends2011,Wenner2013}. Frequencies much above 15 GHz are seldom used because of the difficulty of suppressing parasitic modes, as well as the dramatically higher cost of components and equipment. In addition to the microwave signals, most qubit technologies also have need for relatively slow-changing or even dc signals (typically currents for superconducting qubits; voltages for semiconductor-based qubits), as well as fast (ns-scale) non-sinusoidal pulses for other control operations such as activating tunable couplers to create precise 2-qubit interactions \cite{Chen2014}.

The limited cooling capacities of dilution refrigerators (generally $\ll 1$mW at the 20-30 mK temperatures of interest) and the need to prevent contamination of qubits by any type of non-equilibrium excitations seem to require that the majority of control signals in a quantum computer be generated in a higher-temperature environment, where much greater cooling capacity is available and dissipative circuits will not disrupt nearby qubits. Such signals are then transmitted via shielded impedance-controlled GHz-capable transmission lines to the qubit environment. Depending on the architecture, the number of required transmission lines may someday number into the millions. The use of this many high-performance discrete coaxial cables would seem to be impractical, considering their collective physical size, thermal leakage, and cost. Various multiplexing schemes are being pursued \cite{Hornibrook2014,Chen2012}, which could reduce the number of required lines by $\sim$10$\times$, but the numbers would still be daunting.

The 4 K temperature stage that exists in the upper portion of typical dilution refrigerators provides a particularly attractive location for a layer of control electronics (including basic quantum error correction functions) \cite{Lamb2016,Homulle2016}. Heat dissipation capabilities of multiple Watts are currently available from pulse-tube systems, and it is easy to imagine how the cooling capacity could be further increased using an externally supplied flow cryostat. Large numbers of ultralow-power superconducting electronics can coexist with moderate quantities of semiconductor circuits at 4 K. One can imagine that sufficient electronics could exist at the 4 K stage so that the bandwidth needed to connect to the room-temperature environment is greatly reduced. In this case, the main interconnection problem becomes the links from 4 K down to 20 mK.

\begin{figure}[!hbt]
\centering
\includegraphics[width=6in]{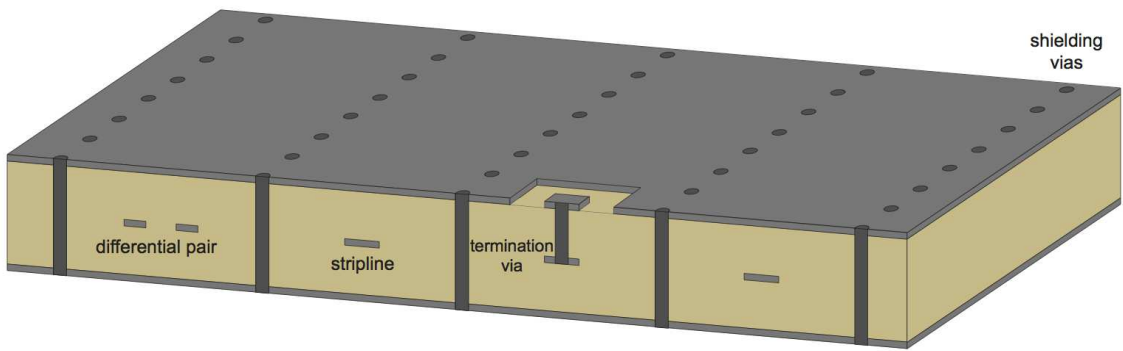}
\caption{Conceptual cross-sectional diagram of a future `rectangular coax' ribbon cable that could be used to communicate between different temperature stages of a dilution refrigerator. The structures reported in this paper are much simpler, but the experimental results provide confidence that much more complex cables, such as illustrated here, could be fabricated and be usable in quantum computing experiments.}
\label{fig:conceptcable}
\end{figure}

We believe that thin-film flexible superconducting transmission lines, fabricated by adapting techniques from the microelectronics industry, can help address this interconnection challenge. Owing to their very low rf surface impedance, superconducting conductors can be scaled to quite small dimensions. Excellent microwave signal propagation can be achieved with conductor cross sections that are orders of magnitude smaller than commercially available coaxial cables. In principle, as illustrated in figure \ref{fig:conceptcable}, one could construct many parallel thin-film transmission lines within a single ribbon cable, with superconducting shield walls (or at least multiple closely spaced vias between upper and lower ground planes) surrounding each signal conductor to form “rectangular coax” cables having negligible crosstalk (an important consideration in many qubit experiments). Typical dimensions of the conductors can be 20 $\mu$m wide and $\sim$ 250 nm thick on a pitch as fine as 50 $\mu$m, which would yield a very high interconnect density (up to 200 signals in a 10 mm wide cable), excellent mechanical flexibility, and very low axial heat leakage when in the superconducting state. Total dielectric (e.g., polyimide) thickness could be 50 $\mu$m or less. Nb ($T_c$ $\sim$ 9.2 K) would be an acceptable superconductor for many applications, although alternative materials could be used if needed for high-magnetic field environments. Either conventional single-ended or differential pairs could be fabricated. Owing to its small cross sectional area, the axial heat leakage along such a ribbon cable would be orders of magnitude less than for an equivalent number of discrete commercially available coaxial cables. The cable terminations could be bonded to a chip using a solder (Pb or In), either in a closely spaced linear array for contact near the edge of a chip, or in a less densely spaced 2-dimensional array, which would mate to the face of a chip or to a connector.

\begin{figure}[!hbt]
\centering
\captionsetup[subfigure]{labelformat=empty}
\includegraphics[width=5in]{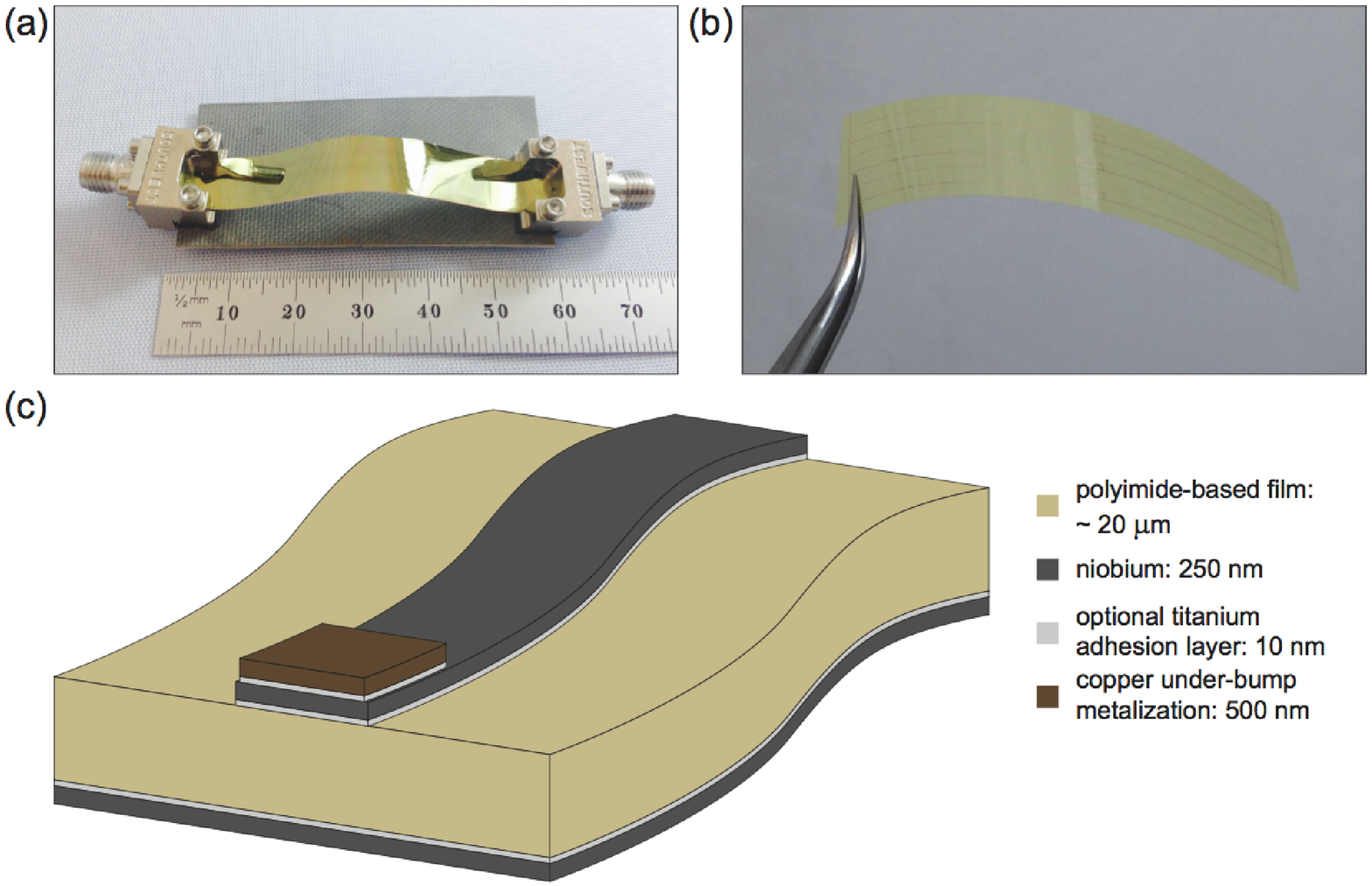}
\subfloat[]{\label{fig:samplea}}
\subfloat[]{\label{fig:sampleb}}
\subfloat[]{\label{fig:samplec}}
\caption{(a) Assembled thin film PI-2611 (meandered) microstrip transmission line with edge launch SMA connectors and support substrate, (b) Released thin film resonator with signal trace, prior to deposition of ground metallization and (c) Cross sectional structure of the simplified structures tested in this work, which were individual superconducting microstrip transmission lines and resonators fabricated on 20-$\mu$m thick polyimide, with lengths ranging from 50 mm to 550 mm. Based on the measured relative dielectric constant of the polyimide at low temperature ($\varepsilon_r$) of 3.2, we chose a line width of 47.4 $\mu$m to yield a 50 $\Omega$ characteristic impedance. Our early test structures were straight 50 mm long lines, and we have since successfully tested serpentine transmission lines and resonators having total lengths of up to 550 mm.}
\label{fig:sample}
\end{figure}

In this work we have fabricated and characterized somewhat simpler structures than proposed in figure \ref{fig:conceptcable}, specifically individual microstrip transmission lines, as a stepping stone towards our ultimate goal. Figure \ref{fig:samplea} is a photograph of one such assembled flexible cable, which is a functional superconducting Nb microwave transmission line built on a thin-film polyimide dielectric.  The performance of these cables, as described herein, gives us confidence that more sophisticated highly parallel cables are indeed feasible.

This paper is organized as follows. In section \ref{prior} we describe prior work and estimate thermal performance of the superconducting thin-film cables. Section \ref{design} presents concepts and details for the design of the microstrip test structures. In section \ref{fab}, we describe relevant fabrication and measurement details. Section \ref{results} covers results for transmission lines, resonators, non-idealities and dielectric loss. Finally, conclusions and closing comments are presented in section \ref{conclusion}.

\section{Relationship to prior work and thermal projections} \label{prior}
A number of researchers have reported making low-frequency (dc to low MHz) superconducting thin film flexible cables  \cite{Yung2011,cheng2013flexible,tighe1999cryogenic,chervenak2013flexible,Weers2013,Bruijn2015}, but none have presented microwave (GHz range) performance data. Manufacturer's data sheets for thin-film polyimides generally only provide low-frequency (kHz or MHz) room-temperature loss tangent (tan$\delta$) data, which is not useful for our purposes.  Some cryogenic loss measurements have been reported for polyimide, but again only at very low frequencies \cite{simon1994properties,weedy1984review,yamaoka1995cryogenic}. Ponchak and Downey \cite{Ponchak1998} have measured room-temperature microwave properties of thin-film copper/polyimide microstrip lines on silicon substrates over the 1-110 GHz frequency range and reported an average tan$\delta$ of 0.006 for PI-2611. Harris et al.\cite{harris2012note} measured copper microstrip lines on free-standing Kapton polyimide and reported a room-temperature tan$\delta$ of 0.013 at 13 GHz, which fell to 0.007 when cooled to 77 K. We will demonstrate in this work that, remarkably, polyimide microwave dielectric losses are reduced by roughly two orders of magnitude when cooled to deep cryogenic temperatures (4 K and below).

Using thin-film geometries similar to those in our work, van Weers et al.\cite{Weers2013} have measured the axial thermal conductivity of a released layer of thin-film PI-2611 to be 1.6$\times10^{-3}T^{1.17}$ W/m$\cdot$K over the temperature range 150 mK to 10 K. In the absence of convective or radiative heat transfer, it can be shown by integrating the one-dimensional Fourier law that whenever a cable's axial thermal conductivity has that type of power-law dependence on absolute temperature, i.e., of the form $k=aT^{n}$, then the steady-state heat $\dot{Q}$ conducting along a cable of uniform cross-sectional area $A$ and length $L$ having a ``hot'' end at temperature $T_{hot}$ and a ``cold'' end at $T_{cold}$ will be $\dot{Q}=(k_{hot} T_{hot}-k_{cold} T_{cold})A/[(n+1)L]$, where $k_{hot}$ and $k_{cold}$ are the thermal conductivities at the hot and cold ends. When $T_{hot}/T_{cold}\gg 1$, as is the case in our contemplated applications, this simplifies to $\dot{Q}=k_{hot} T_{hot} A/[(n+1)L]$. We can thus predict that the heat conduction of a 10 mm wide, 20 $\mu$m-thick polyimide ribbon spanning a 150 mm distance from the 4K stage of a dilution refrigerator to the next lower stage would be only 20 nW, which is insignificant. Cables with a ``hot'' end at a lower temperature stage would be even less conductive, e.g. only 0.6 nW of heat leakage from 800 mK to lower temperatures.

In contrast to the flex cables in \cite{Weers2013}, where the thermal conductivity of their Nb traces were deemed negligible based on a 2000:1 polyimide/Nb area ratio, we use ground planes in our devices, for shielding and precise impedance control at microwave frequencies. Our polyimide/Nb ratio for microstrip is in the range between 40:1 and 80:1 depending on the signal density, and so we cannot neglect the Nb; in fact it may be the dominant contributor to axial heat leakage. The reported thermal conductivity of Nb with an RRR of 40 from T = 4 K down to T = 1.5 K ranged from 9 to 3 W/m$\cdot$K, respectively\cite{Koechlin1996}. Lower-temperature measurements of bulk Nb with RRR = 26 have shown that the thermal conductivity drops monotonically starting from $\sim$ 2 W/m$\cdot$K at 0.7 K, and diminishing to $\sim$ 0.002 W/m$\cdot$K at 40 mK, following fairly close to the theoretical $T^3$ dependence (a $T^{2.4}$ dependence looks like a best fit)\cite{Anderson1971}; $k_{Nb}$ $\sim$ 0.028 W/m$\cdot$K at 120 mK using this fit. In contrast to these bulk samples, our Nb thin films have an RRR of $\sim$ 3, so we might expect both electronic and phononic contributions to thermal conductivity to be reduced by an order of magnitude, but to be very conservative we will assume the bulk values here. The heat leaks of greatest concern in a dilution refrigerator are normally those loading the lowest temperature stage (mixing chamber), where one wants to maintain T $\sim$ 20 mK. The next temperature stage would often be at approximately 120 mK, and so we can assume that cabling is thermally tied at that stage. Using the same calculation method as for the polyimide case above, but with the $n = 2.4$ power law, we estimate that the ground plane of a 10 mm wide, 150 mm long microstrip cable would conduct only 16 pW of heat; again, this is quite negligible, particularly since such a cable could contain as many as 100 parallel transmission lines.

\section{Design of microstrip transmission line structures on thin film polyimide} \label{design}

\begin{figure}[!hbt]
\centering
\includegraphics[width=4in]{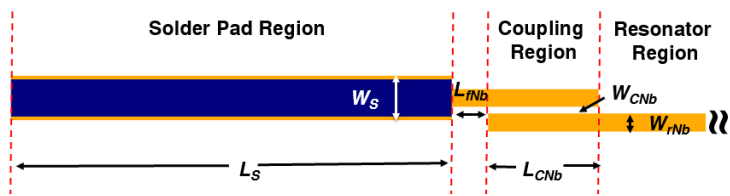}
\caption{Layout of end section of Nb resonator. The length of the resonator region is 46.1 mm, $L_s$ is 1200 $\mu$m, $W_s$ is 120 $\mu$m, $L_{fNb}$ is 100 $\mu$m, $W_{CNb}$ is 20 $\mu$m, $W_{rNb}$ is 47.4 $\mu$m, and $L_{CNb}$ ranges from 300 to 500 $\mu$m.}
\label{fig:layout}
\end{figure}

The structures fabricated and tested in this work are microstrip transmission line elements on thin-film dielectrics that were spin-deposited onto Si substrate wafers and later released to produce free-standing flexible films. Both PI-2611 and HD-4100 polyimides from HD MicroSystems were characterized in this work. In our experience, superconductors deposited on these thin-film dielectrics exhibited higher quality than when we use commercially supplied free-standing Kapton films, because the surface quality is smoother and has fewer defects\cite{RujunIMS2016}. PI-2611, while not photodefineable, is expansion-matched to silicon, exhibits excellent insulation characteristics and high material flexibility, and is widely used as a microwave device packaging and substrate material \cite{meyer2001high}. Photodefineable HD-4100 is also widely used in device packaging since it can be directly patterned using photolithography. Our initial resonator designs were based on the dielectric properties listed on the manufacturer's data sheets, which report PI-2611 to have a relative dielectric constant ($\varepsilon_{r}$) = 2.9 at 1 kHz at room temperature \cite{PI2611}; HD-4100 is reported to have $\varepsilon_r$ = 3.36 at 1 MHz \cite{HD4100}. In order to better determine $\varepsilon_{r}$ at temperatures and frequencies of interest and avoid confounding effects due to superconductor kinetic inductance, which can cause a shift in resonant frequency as the temperature changes, Cu/polyimide resonators were fabricated and their S-parameters were measured at 6.5 K. We used Keysight Advanced Design System (ADS 2015) for simulation of the resonator microwave response and fit the measured data by adjusting the dielectric constant in the simulation. The results indicate the dielectric constant is quite constant over a frequency range from 1 to 10 GHz, and was determined to be $\varepsilon_r$ = 3.2. HD-4100 was assumed to have a similar value for $\varepsilon_r$ and this was later confirmed by superconducting resonator measurements. Based on our fit results, we designed a series of microstrip transmission lines and half-wavelength, series microstrip transmission line resonators using 250 nm thick Nb on 20 $\mu$m thick PI-2611 or HD-4100 films. A linewidth of 47.4 $\mu$m was used to provide a nominal 50 $\Omega$ characteristic impedance. In order to make reliable microwave connections, we used a 120 $\mu$m ($W_s$) $\times$ 1200 $\mu$m solder pad as shown in figure \ref{fig:layout}, which was designed to mate to Southwest Microwave edge launch SMA connectors. The transmission line lengths ranged from 50 mm (straight) to 550 mm (meandered). For the resonators, the solder pads were connected to 47.4 $\mu$m ($W_{rNb}$) $\times$ 100 $\mu$m ($L_{fNb}$) traces as feeding structures at both ends of the resonator. The resonator line length was 46.1 mm, corresponding to a fundamental resonance frequency ($f_0$) of approximately 2 GHz. The resonator coupling structure is shown in figure \ref{fig:layout}. The gap between the coupling lines is fixed at 20 $\mu$m ($W_{CNb}$) and the coupling strength is determined by choosing the overlap length $L_{CNb}$. We simulated and fabricated resonators with different coupling strengths ($L_{CNb}$ = 300, 400, 500 $\mu$m). We note that $L_{CNb}$ = 300 $\mu$m was the weakest coupling in the present designs (an effective series coupling capacitance of 2.76 fF at each end), which turned out to be the preferred choice in view of the unexpectedly low losses of our resonators. 

\section{Fabrication and measurement details} \label{fab}

Si wafers were used as handle wafers and temporary support substrates for fabrication of the thin-film transmission lines and resonators. To start, a Cr/Al layer was deposited onto the Si wafers for use as a release layer later in the fabrication flow. Polyimide (PI-2611 or HD-4100 from HD MicroSystems) was deposited by spin-coating to an approximate thickness of 10 $\mu$m. The films were then cured, using standard curing profiles, in a nitrogen ambient oven at 350 $^{\circ}$C. A second polyimide layer of the same type was then deposited and cured on top of the first layer to achieve a total dielectric thickness of 20 $\mu$m. The resonator traces were defined through a standard photolithographic lift-off process. In our initial samples, we first sputter-deposited a 10 nm thick Ti layer to promote adhesion and potentially improve the Nb film quality, as this had been necessary in earlier work on Kapton films \cite{RujunIMS2016}; we later found this extra Ti layer to be unnecessary when depositing Nb onto thin-film polyimides, and so it was omitted in later samples. Instead, prior to the Nb deposition, a brief Ti getter-pumping step was performed, in which the wafers were stationary behind a shield, while Ti was evaporated into the chamber at a rate of 2 \AA/s for 250 s. After this gettering process, wafers were then rotated at 22 RPM while 250 nm of Nb was sputter-deposited (1 kW dc, 30 minutes, 6.1 mTorr of Ar). Vacuum base pressures of $\sim$ 3 $\times$ 10$^{-7}$ were attained prior to depositing the Nb, which gave fairly repeatable properties. In order to improve the adhesion between the Nb and the polyimide, an in-situ ion milling process was performed for 120 s on the surface of the polyimide before the sputtering process. Si wafer pieces with a SiO$_{2}$ surface were included in the same Nb deposition runs and used as witness samples for characterization of the sheet resistance and superconducting critical transition temperature ($T_{c}$) of the deposited Nb. Typical sheet resistances of our 250 nm thick Nb films ranged from 1 $\Omega$/square to 2.1 $\Omega$/square at room temperature. Typical $T_{c}$ values for the Nb layers used in this work ranged from 8.7 to 9.1 K, with an RRR ($R_{280 K} / R_{10 K}$) between 2.5 and 3.5.

After the Nb deposition, a lift-off process was performed to complete the definition of the signal traces. The samples were then patterned for solder contact pad areas (i.e., under bump metallization, UBM). We used electron beam physical vapor deposition to deposit Ti (50 nm), then Cu (500 nm), onto the samples. The samples were then protected with photoresist and released in a NaCl solution by anodic dissolution \cite{metz2005partial}. A released and cleaned structure is shown in figure \ref{fig:sampleb}. After removing the protective photoresist, we mounted the free-standing film onto another Si support wafer with the non-metallized back side of the film exposed. The sample was then loaded into the deposition system for back-side Nb metallization, with the same thickness as the signal layer. A schematic cross-section of the transmission line structure is shown in figure \ref{fig:samplec}.

In order to facilitate suitable microwave connections to the flexible transmission lines and resonators, we used edge launch SMA connectors. Due to the small dimensions of the signal solder pad, we used Southwest Microwave connectors with the smallest available pin size ($0.005^{\prime\prime}$, i.e. 127 $\mu$m). In order to achieve reliable connections for use at cryogenic temperatures, we soldered the signal pin from the SMA connector to the solder pad on the flex using high purity In solder. Additionally, we prevented the hefty SMA connectors from straining and damaging the thin flexible structures by mounting the connectors and the sample onto a support board, as shown in figure \ref{fig:samplea}.

Measurements above $\sim$ 1 K were carried out in either a LHe dewar (4.2 K), or a Cryo Industries of America pulse-tube based cryostat with stainless steel cryogenic rf coaxial cables. The pulse-tube cryostat with the rf cables stabilizes at approximately 3 K, and can temporarily cool to $\sim$ 1 K by pumping on the sample space that has been back-filled with He. Thermometry was performed at the sample holder near the top of the sample. The sample was primarily thermalized through the He exchange gas. A performance network analyzer (Keysight N5227A PNA) was used for measurement of the scattering parameters. 

The milliKelvin measurements were performed in a 3He/4He dilution refrigerator (Leiden Cryogenics CF450) with base temperature of 20 mK. High frequency coax lines were attenuated at each temperature stage to reduce noise and to thermalize the inner conductors. The Southwest Microwave SMA connectors at each end of the flex cable were thermalized to the mixing chamber via SMA feedthroughs (for the grounds) and cryogenic attenuators (for the center conductors). Transmission measurements were performed using a vector network analyzer (Keysight N5245A), with room temperature amplification (Low Noise Factory LNF-LNR1\_15A) and cryogenic amplification on some samples (Caltech CITLF). Measurement results in the following sections are corrected for system attenuation / amplification, unless otherwise indicated.

In all our measurement set-ups, no intentional magnetic shielding was used, so it is quite possible that flux vortices threading the Nb thin films reduced our resonator $Q$ values below what would be observed in an ideal field-free environment\cite{Song2009}.

\section{Results and discussion} \label{results}

In this section, we describe and discuss results for superconducting microstrip transmission lines and resonators fabricated on thin-film polyimide. These structures are all free-standing films, supported only at the ends by the edge-launch connectors. We also discuss non-ideal low power and high power effects, as well as extraction of upper bounds on tan$\delta$ for the polyimide at various temperatures.

\subsection{Performance of superconducting thin-film polyimide transmission lines}

\begin{figure}[!htb]
\subfloat[]{\includegraphics[height=2.5in]{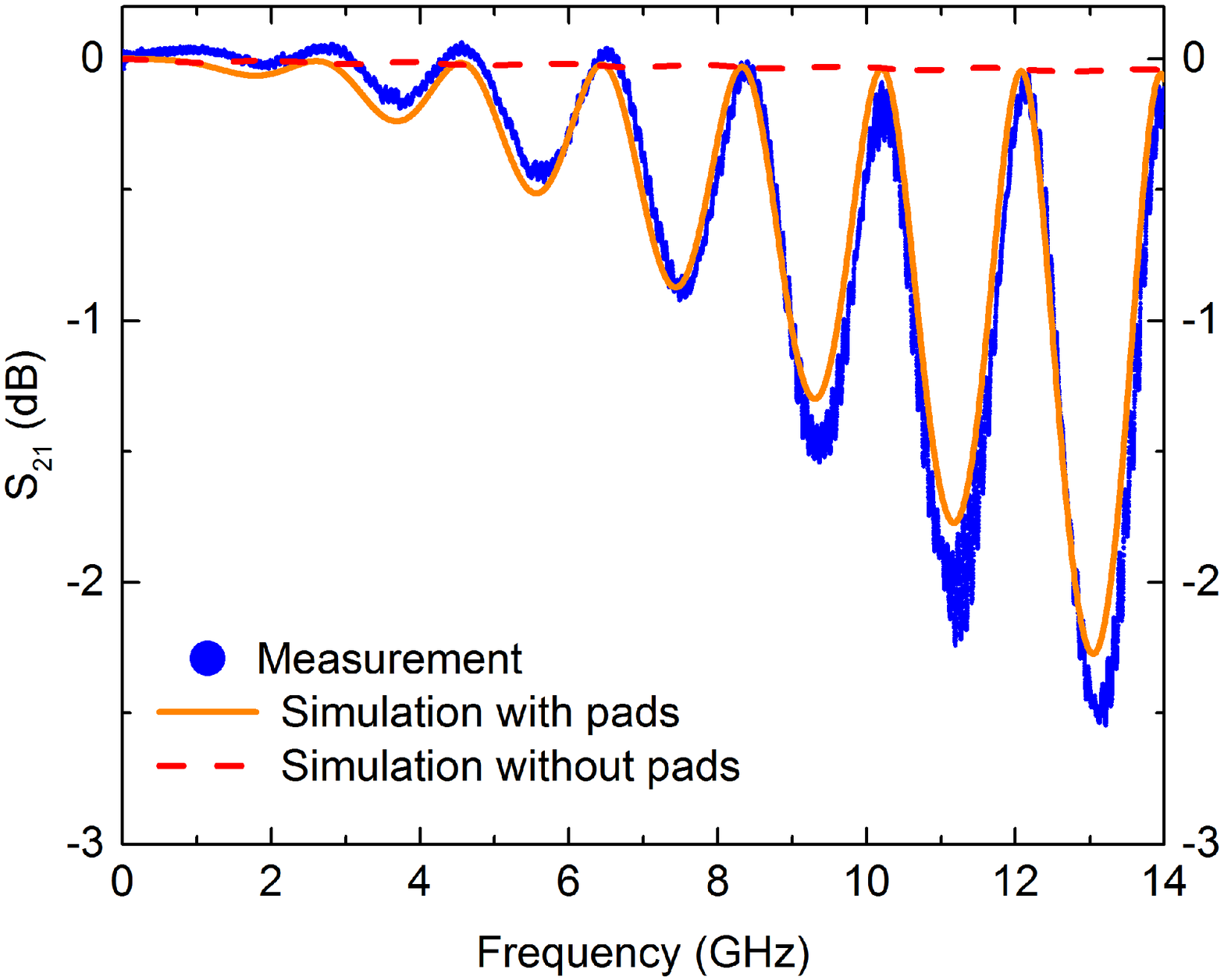}\label{fig:5cmS21}}
\subfloat[]{\includegraphics[height=2.5in]{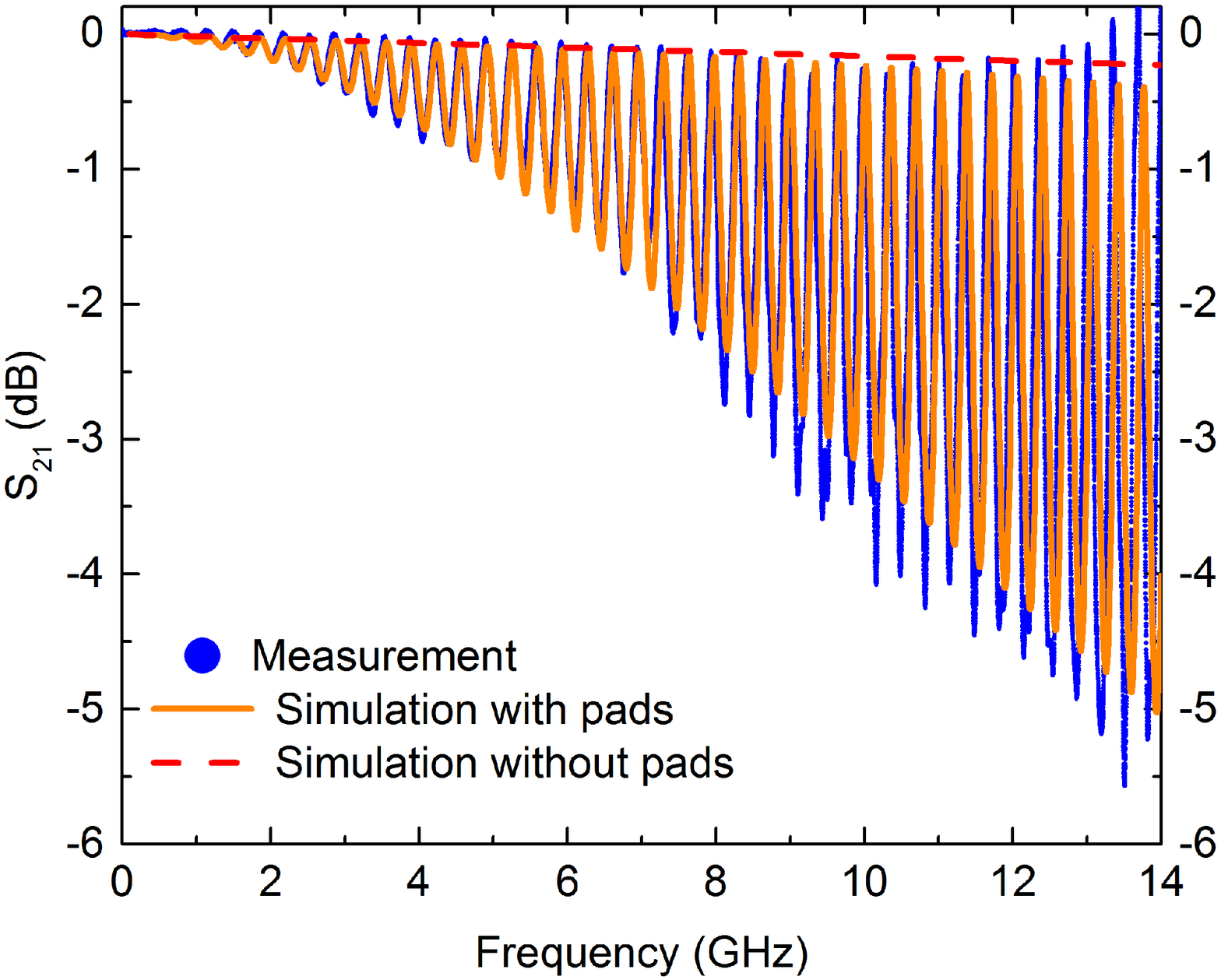}\label{fig:5by5cmS21}}
\caption{Measurement and simulation data of: (a) S$_{21}$ for a 50-mm long flexible superconducting microstrip transmission line (Ti/Nb on PI-2611). Simulation results are shown both with and without the terminal solder pads, to demonstrate the effects of those parasitic capacitive structures. (b) S$_{21}$ for a 5$\times$ longer (i.e., 250 mm), otherwise nominally identical microstrip line.  These samples were immersed in liquid He (T=4.2 K) and tested using -20 dBm incident power; results at higher powers were identical up to 0 dBm, beyond which 3\textsuperscript{rd}-order nonlinearities started to become discernible in the power spectrum.}
\label{fig:sctlineS21}
\end{figure}

S-parameters were measured for superconducting Nb transmission lines at frequencies up to approximately 14 GHz and at an incident microwave power of -20 dbm, while immersed in LHe (4.2 K). SOLR calibration was used, requiring several immersion cycles \cite{AgilentSOLR}. Representative S$_{21}$ results for a 50-mm long line are shown in figure \ref{fig:5cmS21}. The observed oscillations are due to alternating constructive and destructive interference from impedance discontinuities at the two ends of the near-lossless transmission line. Specifically, the connector pin attachment points required a wide soldering pad (see figure \ref{fig:layout}) and additional solder, resulting in parasitic capacitance to ground. To confirm this explanation, Keysight ADS was used to simulate the response of the transmission lines. We used dielectric loss tangent values derived from measurements of superconducting resonators (described below) fabricated on the same substrate as the transmission lines. Excellent matching of simulation to measurement was obtained and is shown as simulation results `with pads'. The variables in these simulations were the kinetic inductance of the superconductor and the two values of the extra capacitance (one at each end of the line, and not necessarily symmetric due to variability in solder quantity). These capacitive pads and accompanying extra solder, which result from the current method of making contact to the structure, are expected to be eliminated with a more refined connection method, such as flip-chip attachment. Therefore, we also performed simulations without these capacitive instances and show these results as simulation `without pads' in figure \ref{fig:5cmS21}; these simulation results closely coincide with the constructively interfering peaks (top envelope) of the S$_{21}$ response, as expected. These results demonstrate that, with ideal signal launchers, insertion losses less than approximately 0.1 dB up to at least 14 GHz can be achieved for a 50-mm long superconducting microstrip transmission line constructed from Nb on flexible polyimide substrates. 

We also fabricated and tested a 5$\times$ longer transmission line (in the form of a 250 mm serpentine). S$_{21}$ for this transmission line is shown in figure \ref{fig:5by5cmS21}, along with a simulation that includes the extra capacitance. Excellent matching between measurement and simulation up to 12 GHz is evident for this structure, as well; the discrepancies above 12 GHz are believed to be imperfections in the calibration procedure. These results show the feasibility of fabricating functional, low-loss flexible superconducting transmission lines with lengths sufficient to span adjacent temperature stages in a typical dilution refrigerator.

The dc critical current ($I_c$) of a representative microstrip signal conductor was measured as 11 mA at 4.2 K.  Although this is well below the critical current density of high-quality Nb films, we have evidence that the reduced $I_c$ is at least partly due to various defects in the Nb film, so we can presumably increase $I_c$ in the future by making process improvements, if it becomes necessary.  $I_c$=11 mA would seem to imply that the transmitted rf power is limited to $\sim$5 dBm, however in the GHz band we were able to transmit at least 10 dBm with negligible loss, although a small amount of nonlinearity in the form of a 3\textsuperscript{rd} harmonic began to appear as the incident power was increased above 0 dBm.  

\subsection{Performance of superconducting thin-film polyimide resonators}

\begin{figure}[!hbt]
\centering
\includegraphics[width=4.5in]{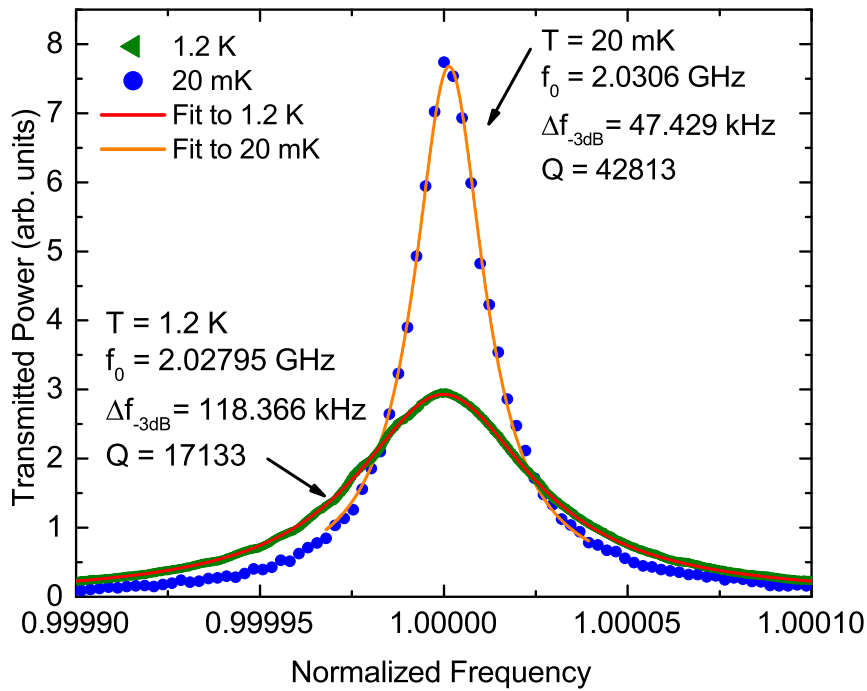}
\caption{Fundamental ($\sim$ 2 GHz) resonance of Nb on HD-4100 microstrip resonator at 1.2 K and 20 mK, showing measured S$_{21}$ and fits to a Lorentzian function. Data has been plotted normalized to center frequency at each temperature. Center frequency, 3dB bandwidth, and resultant loaded $Q$-factor are also provided for each temperature.}
\label{fig:fundamental}
\end{figure}

Due to the low loss of superconducting transmission lines of manageable lengths, detailed characterization of the loss mechanisms is challenging. Instead, we measured the quality factors (``$Q$'') of structurally similar resonators, which provide a sensitive probe of the combined dielectric and superconductor losses \cite{Megrant2012}. For each of the fabricated resonators, S$_{21}$ and S$_{12}$ were measured up to 20 GHz at various temperatures (4.2 K, 3.6 K, 3 K, $\sim$ 1 K and 20 mK). Care was taken to obtain sufficient measurement points in a narrow frequency range around each resonance in order to achieve a reasonable fit to a Lorentzian line shape, which was characteristic of the resonances in these structures and allowed extraction of the loaded $Q$-factor. Figure \ref{fig:fundamental} shows a Nb/HD-4100 resonator with loaded $Q>17$\,$000$ at 1.2 K and $Q>42$\,$800$ at 20 mK. In each of these measurements, the resonance is for a signal power level that provides the highest $Q$. Power levels significantly below this value resulted in lower $Q$ values at 20 mK but not at 1 K or above, while much higher power levels caused various nonlinear effects and distortions of the resonance peak at any temperature. Further details of these effects are described later in this paper.



\begin{figure}[!hbt]
\centering
\includegraphics[width=4.5in]{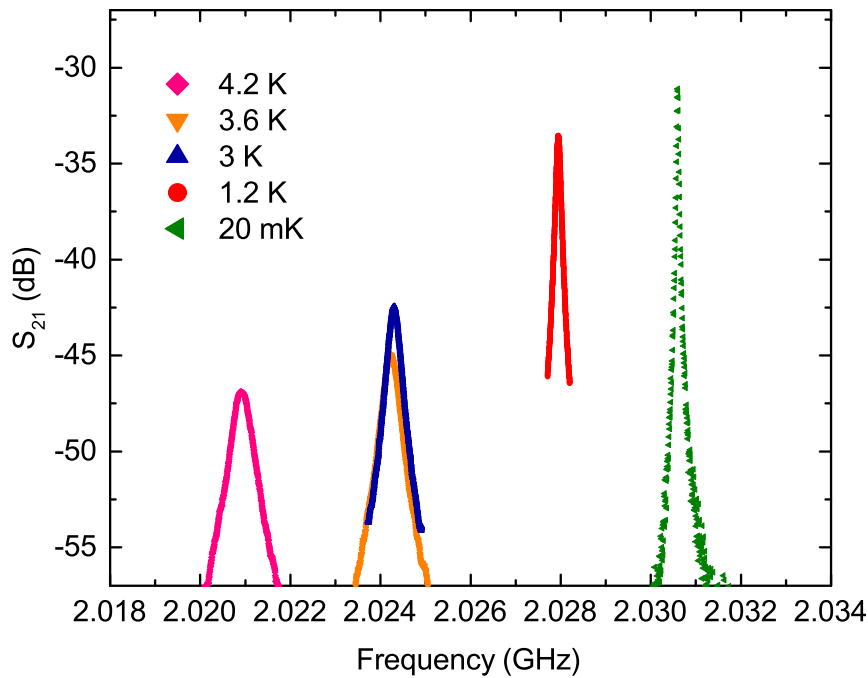}
\caption{$S_{21}$ of fundamental resonance versus frequency at multiple temperatures for a Nb on HD-4100 resonator.}
\label{fig:peakshift}
\end{figure}


In figure \ref{fig:peakshift}, we show the impact of sample temperature on the $Q$-factor and center frequency $f_{0}$ of the fundamental resonance. Reduction of the sample temperature causes a shift in the center frequency towards a higher frequency and an increase in the $Q$-factor. The center frequency shifts can be explained by a reduction of the superconductor kinetic inductance with decrease in sample temperature. Furthermore, as the sample is cooled further below the $T_{c}$, there is a reduction of quasiparticles (un-paired electrons), as expected from BCS theory, which leads to a reduction in attenuation from surface resistance, as well as further reduction in dielectric loss tangent, both of which increase the $Q$-factor. We note that in figure \ref{fig:peakshift}, the 3.6 K data and 3 K data are quite similar. A possible explanation for this is the existence of an interfacial region between the dielectric and Nb superconductor that has different superconducting properties, such as $T_{c}$. Another possibility is that the overlap results from a limitation of the temperature stability of the cryogenic system used for these measurements at temperatures that are not inherently stable in this system. We estimate that the temperature stability of the pulse-tube based cryostat is within 0.5 K at temperatures below 4.2 K.

\begin{figure}[!hbt]
\centering
\captionsetup[subfigure]{labelformat=empty}
\includegraphics[width=5.5in]{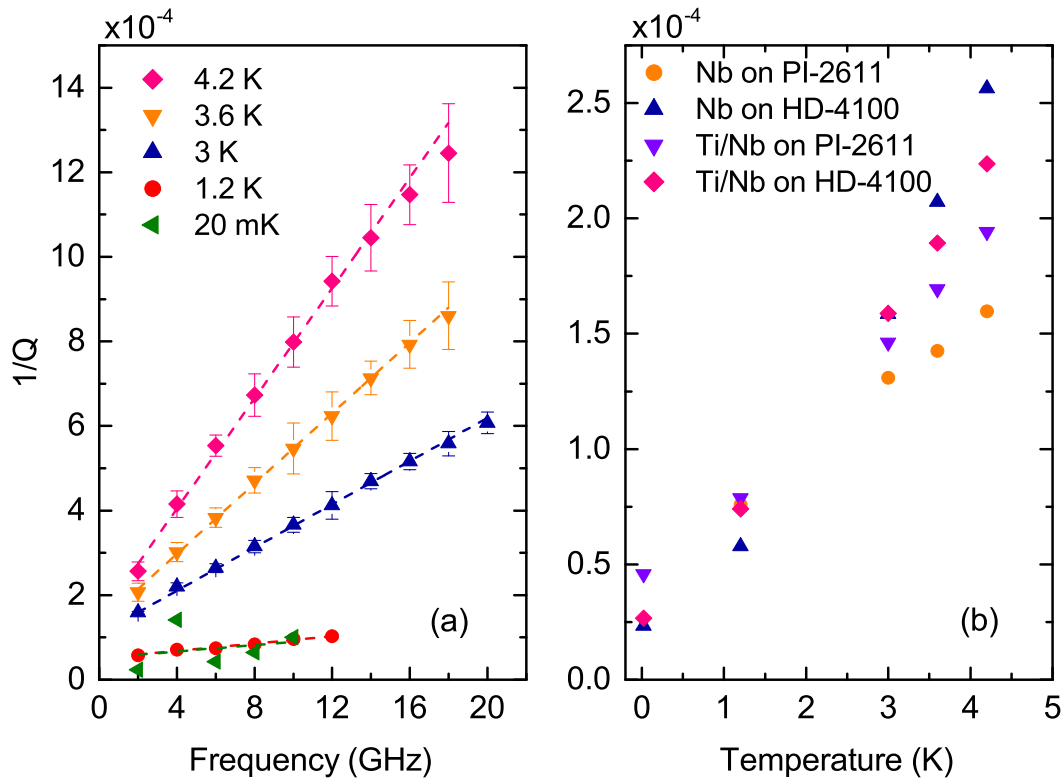}
\subfloat[]{\label{fig:BCSa}}
\subfloat[]{\label{fig:BCSb}}
\caption{(a) 1/$Q$ (loaded) for multiple harmonics of a Nb/HD-4100 microstrip resonator at different temperatures. Dashed lines are linear fits. (b) 1/$Q$ (loaded) as a function of temperature for the fundamental ($\sim$2 GHz) resonance of several PI-2611 and HD-4100 resonators.}
\label{fig:BCS}
\end{figure}

In order to determine the temperature below which superconductor loss becomes negligible, we generated 1/$Q$ versus frequency plots, as shown in figure \ref{fig:BCSa}. In order to consider measurement error and system temperature drift, each of the measured $Q$-factors is the average of ten individual measurement results (for the same sample cool-down) with a 3$\sigma$ error bar. Based on superconducting transmission line resonator theory, a 1/$Q$ versus resonant frequency plot should yield a straight line, with a slope proportional to the magnitude of quasiparticle-induced losses (``BCS losses''). BCS losses actually increase as the square of frequency, but since the $Q$ of a resonator varies inversely with the wavelength, this nets a linear dependence between 1/$Q$ and frequency, which is what we generally observe. (We note that the 2$^{nd}$ harmonic ($\sim$4 GHz) response is sometimes an outlier, showing higher loss than the trend line. This effect has been intermittently observed on some samples, and may be caused by surface contamination of the exposed Nb signal trace, and/or by unintended resonant interactions with the sample chamber. These anomalies are being investigated further.)

The BCS losses should fall exponentially as $T/T_c$ is reduced; this is consistent with figure \ref{fig:BCSa}, which shows a substantial reduction in slope as temperature reduces from 4.2 K to 3 K. The $\sim$1 K data shows a slope of nearly zero; at this temperature, the Nb has minimal BCS loss and the $Q$-factor is presumed to be almost entirely due to dielectric losses, although we cannot rule out the possibility of other residual losses that might be associated with the superconductor, with various interfaces, or with unintended electromagnetic interactions with the environment.

Assuming that the lowest-temperature (1.2 K and 20 mK) $Q$ values are indeed primarily due to polyimide loss, the near-zero slopes in \ref{fig:BCSa} are indicative of a near-constant loss tangent value across the 2-12 GHz frequency range.  This was not unexpected, since $\epsilon_r$ was almost constant at 3.2 across this frequency range (as evidenced by the resonator harmonics occurring at almost exact integer multiples of $f_o$).  Significant changes in loss tangent vs. frequency over a large frequency range normally coincide with noticeable variations in $\epsilon_r$, a consequence of causality (i.e., the Kramers-Kronig relations).

Extrapolating the linear fits in figure \ref{fig:BCSa} to zero frequency provides an estimate of actual dielectric loss by removing the BCS loss contribution (provided that the loss tangent is in fact nearly frequency independent). It is evident that this intercept point falls with decreasing temperature, suggesting that the loss tangent is continuing to fall even at these extremely low temperatures. To provide additional insight into the temperature dependence of the dielectric loss of the polyimide, figure \ref{fig:BCSb} plots 1/$Q$ for the fundamental frequency as a function of temperature for each type of resonator. This data suggests an approximately linear relationship between dielectric loss and absolute temperature, although again we cannot rule out the possibility that the Nb superconductors are also contributing some residual loss, even though the basic BCS theory predicts negligible losses at and below 1.2 K for Nb.


\subsection{Impact of signal power on Q-factor}

\begin{figure}[!hbt]
\centering
\includegraphics[width=4.5in]{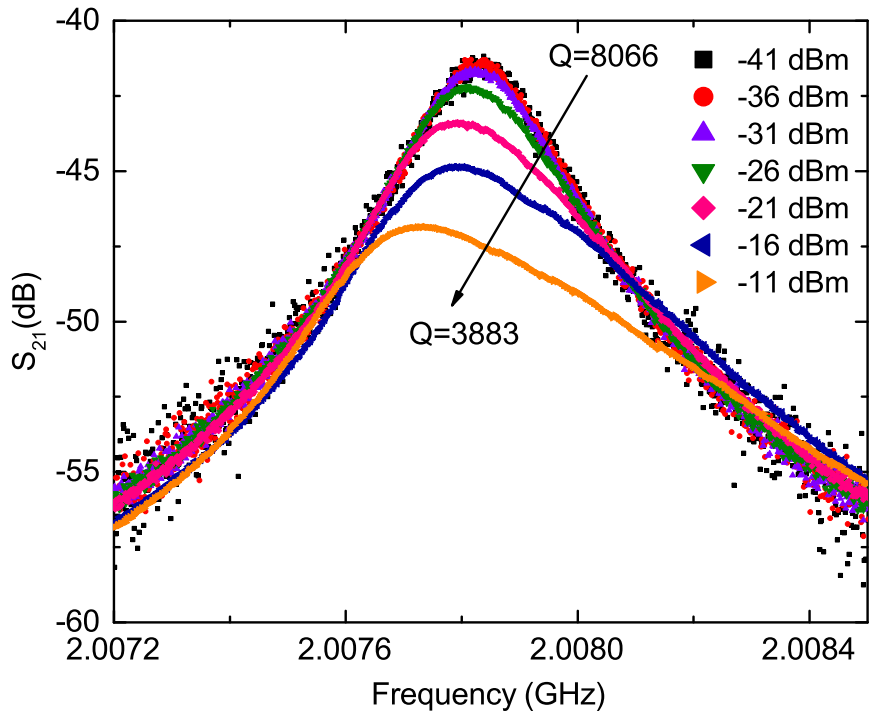}
\caption{Fundamental harmonic response of Nb on PI-2611 resonator at 3 K for different power levels incident at the sample. The range of loaded $Q$ values is shown. Due to the obvious nonlinearity at higher powers, $Q$ was calculated from the measured 3-dB bandwidths, rather than fitting to a Lorentzian function.}
\label{fig:power1}
\end{figure}

\begin{figure}[!hbt]
\centering
\captionsetup[subfigure]{labelformat=empty}
\includegraphics[width=6in]{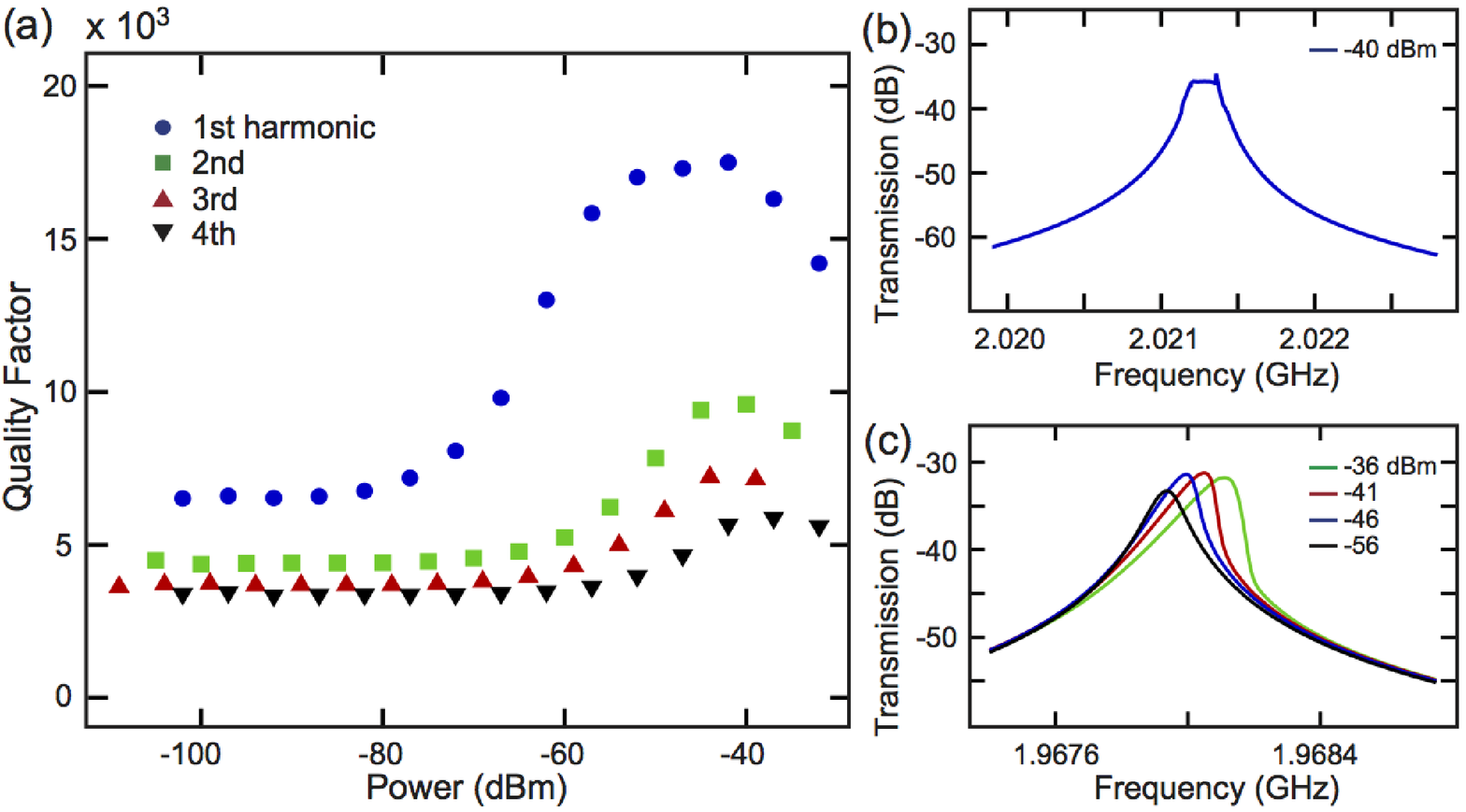}
\subfloat[]{\label{fig:powera}}
\subfloat[]{\label{fig:powerb}}
\subfloat[]{\label{fig:powerc}}
\caption{(a) Loaded $Q$ versus incident power at the sample for the first four harmonics (2-8 GHz) of a Ti/Nb on PI-2611 resonator at 20 mK, (b) fundamental resonance showing high power effects of high $Q$-factor and resonance peak clipping, and (c) fundamental frequency resonances showing Duffing-type non-linearity due to high power effects.}
\label{fig:power}
\end{figure}

 $Q$-factors of these resonators are power dependent, as shown in figures \ref{fig:power1} and \ref{fig:power}. These effects differ significantly depending on the temperature regime.  Figure \ref{fig:power1} illustrates typical behavior at 3 K. As the incident power is increased above -25 dBm, the resonant frequency trends lower, the $Q$ decreases, and the resonance exhibits nonlinearity. This can be explained by the current concentration at the edges of the Nb signal trace, which causes locally increased kinetic inductance and ultimately loss of superconductivity. For the data in figure \ref{fig:BCS}, we used drive power levels just below the onset of this non-linearity.
 
 Figure \ref{fig:power} illustrates representative behavior observed at 20 mK for our resonators. Again, high powers induce nonlinear behavior that reduces $Q$, although the shapes of the resonances suggest that different loss mechanisms are operative at these low temperatures. For example figure \ref{fig:powerb} shows a flat-top profile that may indicate switching of a weak link within the resonator when the local rf current exceeds a critical value. (We observed that after driving the resonator normal with high-power rf and then removing the high power, the resonator would return to its high-$Q$ value after about 5 seconds, indicating that the thermal anchoring was effective.) Figure \ref{fig:powerc} suggests a Kerr nonlinearity associated with partially saturated two level states \cite{suh2013optomechanical}. We also observe a new effect in figure \ref{fig:power} that was not observed at higher temperatures, which is a reduction in $Q$ at very low power levels; this reduction levels off at sufficiently low power. This behavior is likely due to unsaturated two level systems in the polyimide and/or surface oxides on the Nb, which also occur in amorphous dielectric materials such as SiO$_2$ in this milliKelvin temperature regime\cite{OConnell2008,Gao2008,Wang2009,Wenner2011,Neill2003}.
 
\begin{figure}[!hbt]
\centering
\captionsetup[subfigure]{labelformat=empty}
\includegraphics[width=6in]{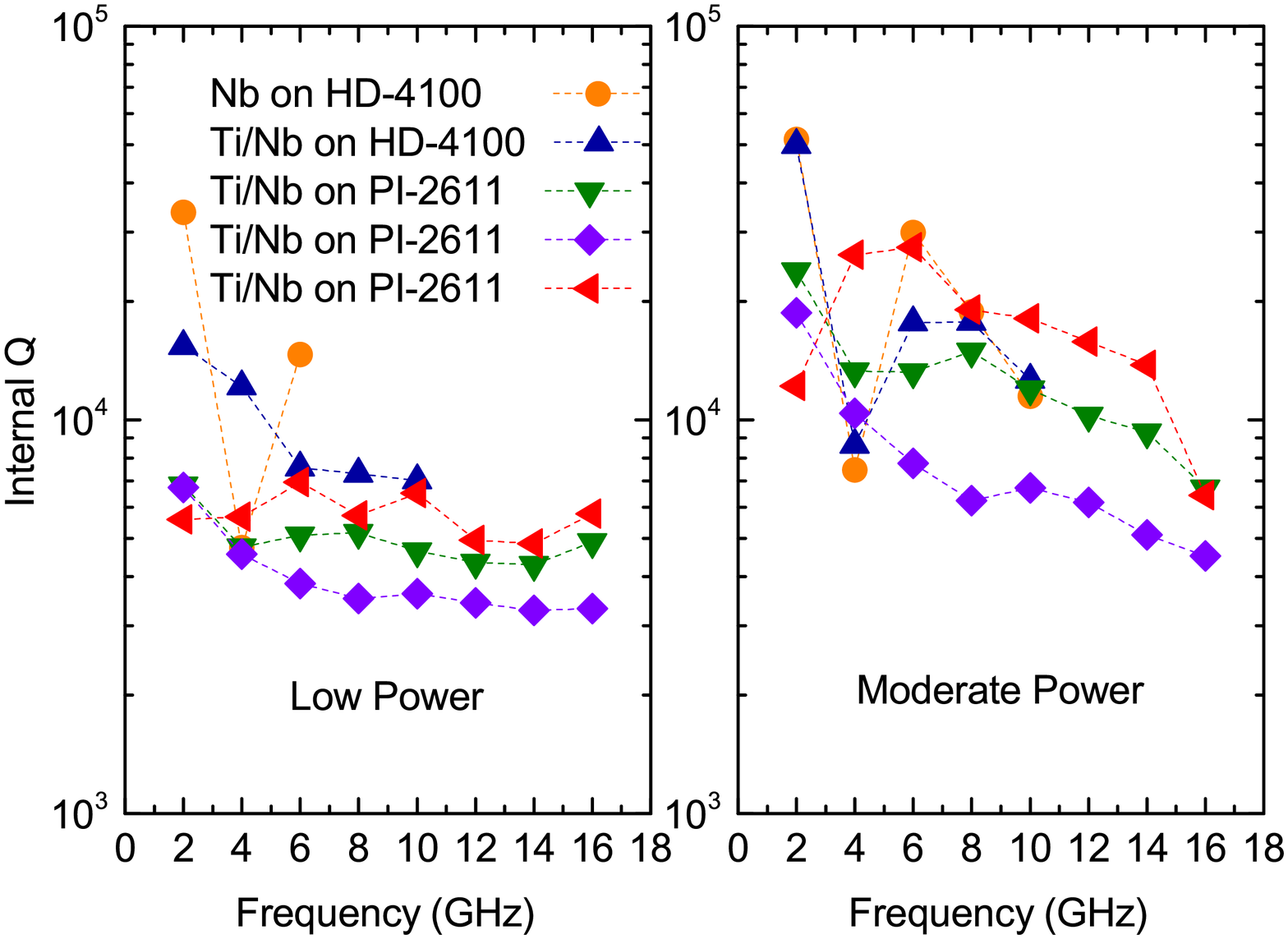}
\subfloat[]{\label{fig:freq_pow_20mKa}}
\subfloat[]{\label{fig:freq_pow_20mKb}}
\caption{Internal $Q$-factors (corrected for coupling loss) versus frequency for PI-2611 and HD-4100 resonators at 20 mK measured in the (a) low power regime and (b) moderate power regime. Results shown here are for five different samples, with the Ti/Nb on PI-2611 being nominally identical.}
\label{fig:freq_pow_20mK}
\end{figure}

 To compare the different materials tested in this work, we plotted the high power and low power $Q$ values at each harmonic to get an idea of the frequency-dependence of the loss tangent. In figure \ref{fig:freq_pow_20mKa} we plot the asymptotic low-power $Q$ values, and in figure \ref{fig:freq_pow_20mKb} we plot the moderate-power $Q$ values, i.e., just below the onset of nonlinearity. We note that above 10 GHz, the incident power was limited by cabling losses and attenuators in the dilution refrigerator, so we were still partially in the low-power regime, which artificially depresses those $Q$ values.

\subsection{Dielectric loss tangent extraction}
To allow more precise modeling and performance predictions for the flexible superconducting transmission lines, knowledge of the dielectric tan$\delta$ for the various polyimide materials is needed. As shown previously, we extracted $Q$-factors at multiple resonant frequencies and multiple temperatures. The extracted $Q$ is the loaded quality factor ($Q_{l}$) and can be described by:



\begin{equation}
\centering
\frac{1}{Q_{l}} = \frac{1}{Q_c}+\frac{1}{Q_d}+\frac{1}{Q_r}+\frac{1}{Q_{coup}}
\label{eqn:1}
\end{equation}

\noindent{}where $Q_c$, $Q_d$, $Q_r$ and $Q_{coup}$ are the $Q$-factors associated with conductor loss, dielectric loss, radiation loss and coupling loss, respectively \cite{qloss}. Radiation losses are assumed to be negligible for these resonators. Given the previously presented 1/$Q$ results (see figure \ref{fig:BCSa}), we assume that the conductor (BCS-related) losses are negligible at temperatures below $\sim$ 1.2 K. Furthermore, ADS simulation results for resonators with no conductor loss and tan$\delta$ = 0 provide values for the coupling loss at each resonant frequency. These values are provided in table \ref{table:1}. With this in mind, we mathematically correct for the coupling loss to get $Q_{d}$ according to:

\begin{equation}
\centering
\frac{1}{Q_{d}} = \frac{1}{Q_l}-\frac{1}{Q_{coup}}
\label{eqn:2}
\end{equation}

\noindent{}From $Q_{d}$, we can determine tan$\delta$ from:

\begin{equation}
\centering
Q_{d} = {\frac{1}{\mbox{tan}\delta}}(1+{\frac{1-q}{q\varepsilon_r}})
\label{eqn:3}
\end{equation}

\noindent{}where $q$ is the dielectric filling factor, which is not unity since this is a non-embedded microstrip structure. From ADS simulation results, with $\varepsilon_r$ = 3.2 and for the case of the microstrip transmission line stack-up used in this work, we find $q$ = 0.703. Based on \ref{eqn:3}, we determine the tan$\delta$ of these two types of polyimide films at various frequencies and temperatures. The loaded $Q$-factor and corresponding tan$\delta$ (corrected for coupling loss) at $\sim$ 1.2 K (for both types of polyimide) and 20 mK (for HD-4100), are shown in table \ref{table:1}. We emphasize again that the existence of other parasitic losses (i.e., not associated with the polyimide dielectric) cannot be ruled out, and so these calculated loss tangents should be viewed as worst-case upper bounds for the actual dielectric loss tangent. 

\begin{table}[!hbt]
\caption{\label{table:1}Measured loaded $Q$-factor and calculated dielectric loss tangent tan$\delta$ at multiple frequencies at $\sim$1.2 K for PI-2611, and at both 1.2 K and 20 mK for HD-4100. The tan$\delta$ calculation corrected for the coupling $Q$ and the non-unity dielectric filling factor of the microstrip. We assume no other loss mechanisms at these low temperatures, and so the actual loss tangents may be smaller than shown.} 

\lineup
\begin{tabular}{lllll}
\br
&&\centre{3}{Loaded $Q$-factor (tan$\delta$)}\\
\ns
&&\crule{3}\\
$\sim f_0$ (GHz) & $Q_{coup}$ & PI-2611 @ 1.2 K & HD-4100 @ 1.2 K & HD-4100 @ 20 mK\\
\mr
\0\0\0\02  & 251$\,$000 & 13$\,$200 (8.21E-5) & 17$\,$300 (6.07E-5) & 42$\,$800 (2.19E-5) \\
\0\0\0\04  & 139$\,$000 & \07$\,$880 (1.35E-4)       & 14$\,$100 (7.19E-5) & \07$\,$070\0(1.52E-4)      \\
\0\0\0\06  & 106$\,$000  & 10$\,$500 (9.72E-5)  & 13$\,$400 (7.34E-5) & 23$\,$300 (3.78E-5) \\
\0\0\0\08  & \088$\,$400  & \09$\,$170 (1.10E-4)       & 12$\,$000 (8.17E-5) & 15$\,$500 (6.00E-5) \\
\0\0\010 & \077$\,$300  & \09$\,$660 (1.02E-4)       & 10$\,$400 (9.42E-4) & 10$\,$000 (9.83E-5) \\
\0\0\012 & \075$\,$400  & \08$\,$870 (1.12E-4)       & \09$\,$710 (1.01E-4)      &                     \\
\br
\end{tabular}
\end{table}

\section{Conclusion and future work} \label{conclusion}

Flexible superconducting transmission line cables, with small physical and thermal cross sections, can be an enabling technology for future computing technologies, such as quantum computing. Our results so far give us confidence that more sophisticated structures can be built. Future work is planned to characterize crosstalk and create embedded and more highly shielded structures. Crosstalk between parallel microstrip conductors follows an inverse square law and so can be made arbitrarily small by spacing the conductors sufficiently far apart, but stripline allows much closer spacing for the same amount of crosstalk. Shield vias, as illustrated in fig. \ref{fig:conceptcable}, will reduce crosstalk even further and allow maximally dense conductor spacing.  The principal challenge to making stripline is the degradation in the superconducting properties (reduced $T_c$ and $I_c$) of the Nb films caused by subsequent high-temperature curing of polyimide.  We have recently found that reducing the polyimide curing temperature substantially alleviates this problem, and so are quite optimistic about the prospects for building more highly shielded cables.

Using a spiral pattern on 100-mm diameter wafers, we have built 2-mm wide cables which were fully 1 meter long after being released.  The use of 300-mm wafers would offer more options for wider and longer cables.  In the long run, we suggest that these sorts of cables could be most economically fabricated using large glass panel substrates, such as are commonly used to manufacture LCD displays.

Connectors could be integrated or flip-chip attached to the ends of the cables. Similarly, attenuators and thermalization structures could be incorporated in various ways, either as integral thin-film patterns or as discrete chips that are flip-chip bonded to the cables. A large number of transmission lines could be integrated within a single relatively narrow ribbon, with outstanding phase matching owing to the integrated construction.

In addition to our flexible cable applications, the extremely low loss exhibited by polyimide thin films at deep cryogenic temperatures suggests that polyimides may also be viable dielectrics for signal distribution on rigid substrates, e.g., for fabricating quantum integrated circuits.



\ack
We gratefully acknowledge funding and technical guidance from Microsoft Research for this work, including valuable discussions with Burton J. Smith. We thank Y. Cao of AMNSTC for assistance with sample fabrication.

\section*{References}
\bibliographystyle{iopart-num}
\bibliography{SC_Flex_arXiv_06142016}

\providecommand{\newblock}{}
\begin{thebibliography}{10}
\expandafter\ifx\csname url\endcsname\relax
  \def\url#1{{\tt #1}}\fi
\expandafter\ifx\csname urlprefix\endcsname\relax\def\urlprefix{URL }\fi
\providecommand{\eprint}[2][]{\url{#2}}

\bibitem{Barends2014}
Barends R, Kelly J, Megrant A, Veitia A, Sank D, Jeffrey E, White T~C, Mutus J,
  Fowler A~G, Campbell B, Chen Y, Chen Z, Chiaro B, Dunsworth A, Neill C,
  O'Malley P, Roushan P, Vainsencher A, Wenner J, Korotkov a~N, Cleland a~N and
  Martinis J~M 2014 {\em Nature\/} {\bf 508} 500--503 ISSN 0028-0836
  \urlprefix\url{http://www.nature.com/doifinder/10.1038/nature13171}

\bibitem{Reed2012}
Reed M~D, DiCarlo L, Nigg S~E, Sun L, Frunzio L, Girvin S~M and Schoelkopf R~J
  2012 {\em Nature\/} {\bf 482} 382--385 ISSN 0028-0836

\bibitem{Kelly2015}
Kelly J, Barends R, Fowler A~G, Megrant A, Jeffrey E, White T~C, Sank D, Mutus
  J~Y, Campbell B, Chen Y, Chen Z, Chiaro B, Dunsworth A, Hoi I~C, Neill C,
  O/'Malley P~J~J, Quintana C, Roushan P, Vainsencher A, Wenner J, Cleland A~N
  and Martinis J~M 2015 {\em Nature\/} {\bf 519} 66--69 ISSN 0028-0836

\bibitem{Chow2015}
Corcoles A~D, Magesan E, Srinivasan S~J, Cross A~W, Steffen M, Gambetta J~M and
  Chow J~M 2015 {\em Nat Commun\/} {\bf 6}

\bibitem{Ofek2016}
Ofek N, Petrenko A, Heeres R, Reinhold P, Leghtas Z, Vlastakis B, Liu Y,
  Frunzio L, Girvin S~M, Jiang L, Mirrahimi M, Devoret M~H and Schoelkopf R~J
  2016 {\em arXiv:1509.01127[quant-ph]\/}  1--44
  \urlprefix\url{http://arxiv.org/abs/1602.04768}

\bibitem{Fowler2012}
Fowler A~G, Mariantoni M, Martinis J~M and Cleland A~N 2012 {\em Physical
  Review A - Atomic, Molecular, and Optical Physics\/} {\bf 86} ISSN 10502947

\bibitem{Mourik2012}
Mourik V, Zuo K, Frolov S~M, Plissard S~R, Bakkers E~P~a~M and Kouwenhoven L~P
  2012 {\em Science\/} {\bf 336} 1003--1007 ISSN 0036-8075

\bibitem{Albrecht2015}
Albrecht S~M, Higginbotham A~P, Madsen M, Kuemmeth F, Jespersen T~S, Nyg{\aa}rd
  J, Krogstrup P and Marcus C~M 2016 {\em Nature\/} {\bf 531} 206--209 ISSN
  0028-0836 \urlprefix\url{http://dx.doi.org/10.1038/nature17162}

\bibitem{Hyart2013}
Hyart T, {Van Heck} B, Fulga I~C, Burrello M, Akhmerov A~R and Beenakker C~W~J
  2013 {\em Physical Review B - Condensed Matter and Materials Physics\/} {\bf
  88} 1--17 ISSN 10980121

\bibitem{Bravyi2005}
Bravyi S and Kitaev A 2004 {\em Physical Review A\/} {\bf 71} 022316 ISSN
  1050-2947 \urlprefix\url{http://link.aps.org/doi/10.1103/PhysRevA.71.022316}

\bibitem{Brecht2015}
Brecht T, Pfaff W, Wang C, Chu Y, Frunzio L, Devoret M~H and Schoelkopf R~J
  2015 {\em arXiv:1509.01127v1 [quant-ph]\/} ISSN 2056-6387
  \urlprefix\url{http://arxiv.org/abs/1509.01127}

\bibitem{Hornibrook2015}
Hornibrook J~M, Colless J~I, {Conway Lamb} I~D, Pauka S~J, Lu H, Gossard A~C,
  Watson J~D, Gardner G~C, Fallahi S, Manfra M~J and Reilly D~J 2015 {\em
  Physical Review Applied\/} {\bf 3} 024010 ISSN 2331-7019
  \urlprefix\url{http://link.aps.org/doi/10.1103/PhysRevApplied.3.024010}

\bibitem{Chow2013}
Chow J~M, Gambetta J~M, Cross A~W, Merkel S~T, Rigetti C and Steffen M 2013
  {\em New Journal of Physics\/} {\bf 15} 115012 ISSN 1367-2630
  \urlprefix\url{http://stacks.iop.org/1367-2630/15/i=11/a=115012?key=crossref.4197d773e3605093549b2c1ae98d3933}

\bibitem{Leek2009}
Leek P~J, Filipp S, Maurer P, Baur M, Bianchetti R, Fink J~M, G{\"{o}}ppl M,
  Steffen L and Wallraff A 2009 {\em Physical Review B\/} {\bf 79} 180511 ISSN
  1098-0121 \urlprefix\url{http://link.aps.org/doi/10.1103/PhysRevB.79.180511}

\bibitem{Wallraff2005}
Wallraff A, Schuster D~I, Blais A, Frunzio L, Majer J, Devoret M~H, Girvin S~M
  and Schoelkopf R~J 2005 {\em Physical Review Letters\/} {\bf 95} 1--4 ISSN
  00319007

\bibitem{Macklin2015}
Macklin C, O'Brien K, Hover D, Schwartz M~E, Bolkhovsky V, Zhang X, Oliver W~D
  and Siddiqi I 2015 {\em Science\/} {\bf 350} 307--310 ISSN 0036-8075
  \urlprefix\url{http://www.sciencemag.org/cgi/doi/10.1126/science.aaa8525}

\bibitem{Corcoles2011}
Córcoles A~D, Chow J~M, Gambetta J~M, Rigetti C, Rozen J~R, Keefe G~A,
  Beth~Rothwell M, Ketchen M~B and Steffen M 2011 {\em Applied Physics
  Letters\/} {\bf 99} 181906

\bibitem{Barends2011}
Barends R, Wenner J, Lenander M, Chen Y, Bialczak R~C, Kelly J, Lucero E,
  O’Malley P, Mariantoni M, Sank D, Wang H, White T~C, Yin Y, Zhao J, Cleland
  A~N, Martinis J~M and Baselmans J~J~A 2011 {\em Applied Physics Letters\/}
  {\bf 99} 113507

\bibitem{Wenner2013}
Wenner J, Yin Y, Lucero E, Barends R, Chen Y, Chiaro B, Kelly J, Lenander M,
  Mariantoni M, Megrant A, Neill C, O'Malley P~J~J, Sank D, Vainsencher A, Wang
  H, White T~C, Cleland A~N and Martinis J~M 2013 {\em Phys. Rev. Lett.\/} {\bf
  110}(15) 150502

\bibitem{Chen2014}
Chen Y, Neill C, Roushan P, Leung N, Fang M, Barends R, Kelly J, Campbell B,
  Chen Z, Chiaro B, Dunsworth A, Jeffrey E, Megrant A, Mutus J~Y, O'Malley
  P~J~J, Quintana C~M, Sank D, Vainsencher A, Wenner J, White T~C, Geller M~R,
  Cleland A~N and Martinis J~M 2014 {\em Phys. Rev. Lett.\/} {\bf 113}(22)
  220502 \urlprefix\url{http://link.aps.org/doi/10.1103/PhysRevLett.113.220502}

\bibitem{Hornibrook2014}
Hornibrook J~M, Colless J~I, Mahoney A~C, Croot X~G, Blanvillain S, Lu H,
  Gossard A~C and Reilly D~J 2014 {\em Applied Physics Letters\/} {\bf 104}
  103108 ISSN 0003-6951
  \urlprefix\url{http://scitation.aip.org/content/aip/journal/apl/104/10/10.1063/1.4868107}

\bibitem{Chen2012}
Chen Y, Sank D, O'Malley P, White T, Barends R, Chiaro B, Kelly J, Lucero E,
  Mariantoni M, Megrant A, Neill C, Vainsencher A, Wenner J, Yin Y, Cleland A~N
  and Martinis J~M 2012 {\em Applied Physics Letters\/} {\bf 101} 182601 ISSN
  00036951
  \urlprefix\url{http://scitation.aip.org/content/aip/journal/apl/101/18/10.1063/1.4764940}

\bibitem{Lamb2016}
Conway~Lamb I~D, Colless J~I, Hornibrook J~M, Pauka S~J, Waddy S~J, Frechtling
  M~K and Reilly D~J 2016 {\em Review of Scientific Instruments\/} {\bf 87}
  014701

\bibitem{Homulle2016}
Homulle H, Visser S, Patra B, Ferrari G, Prati E, Sebastiano F and Charbon E
  2016 A reconfigurable cryogenic platform for the classical control of
  scalable quantum computers (\textit{Preprint} \eprint{arXiv:1602.05786})

\bibitem{Yung2011}
Yung C~S and Moeckly B~H 2011 {\em IEEE Transactions on Applied
  Superconductivity\/} {\bf 21} 107--110 ISSN 10518223

\bibitem{cheng2013flexible}
Cheng M~Y, Park W~T, Yu A, Xue R~F, Tan K~L, Yu D, Lee S~H, Gan C~L and Je M
  2013 {\em Microsystem technologies\/} {\bf 19} 1111--1118

\bibitem{tighe1999cryogenic}
Tighe T, Akerling G and Smith A 1999 {\em Applied Superconductivity, IEEE
  Transactions on\/} {\bf 9} 3173--3176

\bibitem{chervenak2013flexible}
Chervenak J and Mateo J 2013 {\em NASA Technical Brief\/}  5--6

\bibitem{Weers2013}
van Weers H, Kunkel G, Lindeman M and Leeman M 2013 {\em Cryogenics\/} {\bf
  55-56} 1--4 ISSN 00112275
  \urlprefix\url{http://linkinghub.elsevier.com/retrieve/pii/S0011227512002068}

\bibitem{Bruijn2015}
Bruijn M~P, van~der Linden A~J, Ridder M~L and van Weers H~J 2015 {\em Journal
  of Low Temperature Physics\/} ISSN 0022-2291
  \urlprefix\url{http://link.springer.com/10.1007/s10909-015-1360-4}

\bibitem{simon1994properties}
Simon N 1994 {\em NIST Internal/Interagency Reports 5030\/}  1--205

\bibitem{weedy1984review}
Weedy B and Swingler S 1984 {\em Cryogenics\/} {\bf 24} 367--370

\bibitem{yamaoka1995cryogenic}
Yamaoka H, Miyata K and Yano O 1995 {\em Cryogenics\/} {\bf 35} 787--789

\bibitem{Ponchak1998}
Ponchak G~E and Downey A~N 1998 {\em IEEE Transactions on Components,
  Packaging, and Manufacturing Technology: Part B\/} {\bf 21} 171--176 ISSN
  1070-9894

\bibitem{harris2012note}
Harris A, Sieth M, Lau J, Church S, Samoska L and Cleary K 2012 {\em Review of
  Scientific Instruments\/} {\bf 83} 086105

\bibitem{Koechlin1996}
Koechlin F and Bonin B 1999 {\em Superconductor Science and Technology\/} {\bf
  9} 453--460 ISSN 0953-2048

\bibitem{Anderson1971}
Anderson A~C, Satterthwaite C~B and Smith S~C 1971 {\em Physical Review B\/}
  {\bf 3} 3762--3764 ISSN 0556-2805
  \urlprefix\url{http://link.aps.org/doi/10.1103/PhysRevB.3.3762}

\bibitem{RujunIMS2016}
Bai R, Hernandez G~A, Cao Y, Sellers A~J, Ellis C~D, Tuckerman D~B and Hamilton
  M~C 2016 Cryogenic microwave characterization of kapton polyimide using
  superconducting resonators {\em International Microwave Symposium (IMS)\/}
  (San Francisco)

\bibitem{meyer2001high}
Meyer J~U, Stieglitz T, Scholz O, Haberer W and Beutel H 2001 {\em Advanced
  Packaging, IEEE Transactions on\/} {\bf 24} 366--374

\bibitem{PI2611}
HD MicroSystems 2009 {\em PI-2600 Series-Low Stress Applications\/} product
  bulletin

\bibitem{HD4100}
HD MicroSystems 2014 {\em HD-4100 Series\/} product bulletin

\bibitem{metz2005partial}
Metz S, Bertsch A and Renaud P 2005 {\em Microelectromechanical Systems,
  Journal of\/} {\bf 14} 383--391

\bibitem{Song2009}
Song C, Heitmann T~W, DeFeo M~P, Yu K, McDermott R, Neeley M, Martinis J~M and
  Plourde B~L~T 2009 {\em Phys. Rev. B\/} {\bf 79}(17) 174512
  \urlprefix\url{http://link.aps.org/doi/10.1103/PhysRevB.79.174512}

\bibitem{AgilentSOLR}
Ferrero A and Pisani U 1992 {\em IEEE Microwave and Guided Wave Letters\/} {\bf
  2} 505--507 ISSN 1051-8207

\bibitem{Megrant2012}
Megrant A, Neill C, Barends R, Chiaro B, Chen Y, Feigl L, Kelly J, Lucero E,
  Mariantoni M, O’Malley P~J~J, Sank D, Vainsencher A, Wenner J, White T~C,
  Yin Y, Zhao J, Palmstrøm C~J, Martinis J~M and Cleland A~N 2012 {\em Applied
  Physics Letters\/} {\bf 100} 113510

\bibitem{suh2013optomechanical}
Suh J, Weinstein A and Schwab K 2013 {\em Applied Physics Letters\/} {\bf 103}
  052604
  \urlprefix\url{http://www.sciencedirect.com/science/article/pii/0011227584900924}

\bibitem{OConnell2008}
O’Connell A~D, Ansmann M, Bialczak R~C, Hofheinz M, Katz N, Lucero E,
  McKenney C, Neeley M, Wang H, Weig E~M, Cleland A~N and Martinis J~M 2008
  {\em Applied Physics Letters\/} {\bf 92} 112903
  \urlprefix\url{http://scitation.aip.org/content/aip/journal/apl/92/11/10.1063/1.2898887}

\bibitem{Gao2008}
Gao J, Daal M, Martinis J~M, Vayonakis A, Zmuidzinas J, Sadoulet B, Mazin B~A,
  Day P~K and Leduc H~G 2008 {\em Applied Physics Letters\/} {\bf 92} 212504

\bibitem{Wang2009}
Wang H, Hofheinz M, Wenner J, Ansmann M, Bialczak R~C, Lenander M, Lucero E,
  Neeley M, O’Connell A~D, Sank D, Weides M, Cleland A~N and Martinis J~M
  2009 {\em Applied Physics Letters\/} {\bf 95} 233508

\bibitem{Wenner2011}
Wenner J, Barends R, Bialczak R~C, Chen Y, Kelly J, Lucero E, Mariantoni M,
  Megrant A, O’Malley P~J~J, Sank D, Vainsencher A, Wang H, White T~C, Yin Y,
  Zhao J, Cleland A~N and Martinis J~M 2011 {\em Applied Physics Letters\/}
  {\bf 99} 113513

\bibitem{Neill2003}
Neill C, Megrant A, Barends R, Chen Y, Chiaro B, Kelly J, Mutus J~Y, O'Malley
  P~J~J, Sank D, Wenner J, White T~C, Yin Y, Cleland A~N and Martinis J~M 2013
  {\em Applied Physics Letters\/} {\bf 103} 072601

\bibitem{qloss}
Belohoubek E and Denlinger E 1975 {\em Microwave Theory and Techniques, IEEE
  Transactions on\/} {\bf 23} 522--526 ISSN 0018-9480

\end{thebibliography}
\end{document}